\documentclass[submission, Phys]{SciPost}
\pdfoutput=1
\usepackage{amsmath,amssymb,mathtools,xspace}
\usepackage{booktabs,multirow,graphicx,tabularx,slashed}
\usepackage{hyperref}
\usepackage{color,xcolor}
\usepackage[normalem]{ulem}
\usepackage{enumitem}
\usepackage{braket}
\usepackage{stackrel}
\usepackage{tikz}
\usepackage{float}
\usepackage{subcaption}

\graphicspath{{plots/}}

\makeatletter
\def\BState{\State\hskip-\ALG@thistlm}
\makeatother

\makeatletter
\@ifundefined{pdfoutput}{}{\DeclareGraphicsRule{*}{mps}{*}{}}
\makeatother

\makeatletter
\DeclareRobustCommand*{\bfseries}{%
   \not@math@alphabet\bfseries\mathbf
   \fontseries\bfdefault\selectfont
   \boldmath
}
\makeatother

\setitemize{itemsep=2pt,topsep=2pt,parsep=0pt,partopsep=0pt,leftmargin=*}
\setenumerate{itemsep=0pt,topsep=2pt,parsep=0pt,partopsep=0pt,labelindent=10pt,leftmargin=*}
\definecolor{Gcolor}{HTML}{3b528b}
\definecolor{Dcolor}{HTML}{e41a1c}

\tikzstyle{generator} = [rectangle, rounded corners, minimum width=3cm, minimum height=1cm,text centered, draw=Gcolor]
\tikzstyle{discriminator} = [rectangle, rounded corners, minimum width=3cm, minimum height=1cm,text centered, draw=Dcolor]
\tikzstyle{io} = [circle, trapezium left angle=70, trapezium right angle=110, minimum width=1cm, minimum height=1cm, text centered, draw=black]

\tikzstyle{process} = [rectangle, minimum width=1cm, minimum height=1cm, text centered, draw=black]
\tikzstyle{decision} = [rectangle, minimum width=1cm, minimum height=1cm, text centered, draw=black]

\tikzstyle{arrow} = [thick,->,>=stealth]
\usepackage{xcolor}

\newcommand{\XLangle}{\Big\langle}
\newcommand{\XRangle}{\Big\rangle}

\newcommand\one{\leavevmode\hbox{\small1\normalsize\kern-.33em1}}

\newcommand{\loss}{\mathcal{L}}

\newcommand{\qqquad}{\qquad \qquad}
\newcommand{\qqqquad}{\qquad \qquad \qquad}

\newcommand{\gev}{\text{GeV}}

\def\slashchar#1{\setbox0=\hbox{$#1$}           
   \dimen0=\wd0                                 
   \setbox1=\hbox{/} \dimen1=\wd1               
   \ifdim\dimen0>\dimen1                        
      \rlap{\hbox to \dimen0{\hfil/\hfil}}      
      #1                                        
   \else                                        
      \rlap{\hbox to \dimen1{\hfil$#1$\hfil}}   
      /                                         
   \fi}

\newcommand{\ie}{\textsl{i.e.}\;}

\begin{document}

\begin{center}{\Large \textbf{
      Symmetries, Safety, and Self-Supervision 
}}\end{center}

\begin{center}
Barry M. Dillon\textsuperscript{1}, 
Gregor Kasieczka\textsuperscript{2}, 
Hans Olischlager\textsuperscript{1},
Tilman Plehn\textsuperscript{1}, \\
Peter Sorrenson\textsuperscript{3}, and
Lorenz Vogel\textsuperscript{1}
\end{center}

\begin{center}
{\bf 1} Institut f\"ur Theoretische Physik, Universit\"at Heidelberg, Germany\\
{\bf 2} Institut f\"ur Experimentalphysik, Universit\"at Hamburg, Germany \\
{\bf 3} Heidelberg Collaboratory for Image Processing, Universit\"at Heidelberg, Germany
\end{center}

\begin{center}
\today
\end{center}

\section*{Abstract}
{\bf Collider searches face the challenge of defining a representation
  of high-dimensional data such that physical symmetries are manifest,
  the discriminating features are retained, and the choice of
  representation is new-physics agnostic. We introduce JetCLR to solve
  the mapping from low-level data to optimized observables though
  self-supervised contrastive learning.  As an example, we construct a
  data representation for top and QCD jets using a
  permutation-invariant transformer-encoder network and visualize its
  symmetry properties. We compare the JetCLR representation with
  alternative representations using linear classifier tests
  and find it to work quite well.}

\vspace{10pt}
\noindent\rule{\textwidth}{1pt}
\tableofcontents\thispagestyle{fancy}
\noindent\rule{\textwidth}{1pt}
\vspace{10pt}

\newpage
\section{Introduction}
\label{sec:intro}

Symmetries~\cite{Noether1918} form the core of the fundamental
description, phenomenological techniques, and experimental analyses in
particle physics. LHC physics is defined by the symmetry structure of
LHC data, from the detector geometry to the relativistic space-time
symmetries and local gauge symmetries defining the underlying theory,
and to new physics motivations like supersymmetry. Any new approach to
LHC physics, including applications of machine learning, has to be
seen in the context of symmetries
eventually~\cite{Krippendorf:2020gny,Barenboim:2021vzh,Maiti:2021fpy,Krippendorf:2021uxu}.

Modern machine learning (ML) has spurred the development of techniques
which can, among other benefits, boost the development of high-level
observables. We typically train a neural network to distinguish
between different physical processes either based on low-level,
high-dimensional data or on the corresponding Monte-Carlo
simulations. The resulting classifier can be viewed as a high-level
observable for a given analysis. If the dataset can be understood
through first-principle simulations and is large enough to train the
networks, this observable will be optimized for the respective task,
but lack theoretical calculability. As long as the observable is
calibrated and the systematic errors are understood, this lack of
calculability is not a barrier for supervised classification.

This agnostic approach works well for supervised analyses, but it is
not clear how it can be expanded to unsupervised analyses, like an
anomaly
search~\cite{Heimel:2018mkt,Farina:2018fyg,Nachman:2020lpy,Bortolato:2021zic,Dillon:2021nxw},
a generalized side band
analysis~\cite{Metodiev:2017vrx,Kasieczka:2020pil}, or a generalized
model hypothesis~\cite{Barron:2021btf}.  For this purpose, we propose
to replace a limited number of high-level observables by a
high-dimensional representation, and replace full control over all
possible physics processes with a structure driven by symmetries and
fundamental theory.

The standard application driving ML-methods in LHC physics is jets, a
fertile ground for supervised and unsupervised techniques. The most
common jet representation is jet
images~\cite{Cogan:2014oua,deOliveira:2015xxd,Kasieczka:2017nvn,Lin:2018cin,Komiske:2016rsd,Macaluso:2018tck},
a high-dimensional representation defined in rapidity vs azimuthal
angle, observables inspired by Lorentz
transformations. Jet images typically include a pre-processing step
exploiting their rotation symmetry.  Alternative symmetry-inspired
jet representations include permutation-invariant
graphs~\cite{Henrion:DLPS2017,Qasim:2019otl,Chakraborty:2019imr,1808887},
trees~\cite{Louppe:2017ipp,Andreassen:2018apy,Dillon:2019cqt,Dillon:2020quc},
the Lund plane~\cite{Dreyer_2018,Carrazza:2019cnt}, Lorentz-inspired networks~\cite{Butter:2017cot,Erdmann:2018shi,Bogatskiy:2020tje,1811770,Shimmin:2021pkm}, or Energy Flow Polynomials
(EFPs)~\cite{Komiske:2017aww}, a calculable basis with a notion of
infrared and collinear safety.

Combining unsupervised learning and symmetries we define jet
observables using Contrastive Learning of Representations
(CLR)~\cite{49372}. Our key idea is to frame the mapping between the
jet constituents' phase space and a representation space as an
optimization task with a contrastive loss function, \textsl{designed
  such that the representation space will be invariant to pre-defined
  symmetries and retains discriminative power.}  The training is
self-supervised in view of the network's discriminative power, because
the optimization never uses truth labels for the jets.  For the
mapping of physics and representation spaces we employ a
transformer-encoder
network~\cite{Mikuni:2020wpr,Mikuni:2021pou,Shmakov:2021qdz}. In
addition to its built-in permutation symmetry we implement rotation
and translation symmetries, as well as soft and collinear safety
augmentations. To benchmark JetCLR we use a standard test in the ML
community, the so-called linear classifier test (LCT). For this test a
linear network is trained to classify between different processes,
quantifying how well classes can be separated by a linear cut in
representation space.

We start by introducing contrastive learning in Sec.~\ref{sec:clr}. We
then construct our JetCLR tool using a set of symmetries and
augmentations in Sec.~\ref{sec:jetclr}. In Sec.~\ref{sec:results} we
visualize the invariances of the JetCLR representation and study its
performance using a linear classifier. Different such classifiers are
discussed in the Appendix.

\section{Contrastive learning}
\label{sec:clr}

The goal of our network is to define a mapping between the jet
constituents and a representation space,
\begin{align}
  f: \; \mathcal{J} \rightarrow \mathcal{R} \;,
\label{eq:mapping}
\end{align}
which is, both,
\begin{enumerate}
\item invariant to symmetries and theory-driven augmentations, and
\item discriminative within the dataset it is optimized on.
\end{enumerate}
We work with the top-tagging
dataset~\cite{Heimel:2018mkt,Kasieczka:2019dbj,benato2021shared},
where the jets are simulated with Pythia8 \cite{Sjostrand:2014zea}
(default tune) using a center-of-mass energy of 14~TeV and ignoring
pile-up and multi-parton interactions.
Delphes~\cite{deFavereau:2013fsa} provides a fast detector simulation
with the default ATLAS detector card, and the jets are defined through
anti-$k_T$ algorithm~\cite{Cacciari:2008gp,Cacciari:2005hq} in
FastJet~\cite{Cacciari:2011ma} with a radius of $R = 0.8$.  For each
event we keep the leading jet, provided
\begin{align}
  p_T = 550~...~650~\gev
  \qquad \text{and} \qquad
  |\eta | < 2 \; .
\end{align}
This narrow $p_T$-range induces the most distinctive feature in the
jets, a finite geometric distance between the top decay products in
the $\eta - \phi$ plane, whereas for QCD jets the average activity
continuously drops away from the hardest constituent.  The top jets
are required to be matched to a parton-level top and all parton-level
decay products to lay within the jet radius. The jet constituents are
defined using the Delphes energy-flow algorithm, with the leading 200
constituents from each jet kept for the analysis. Particle-ID and
tracking information are not included.

If we assume all jet constituents to be massless, each jet $x_i$ is defined
by its constituent coordinates,
\begin{align}
  x_i = \{ \left( p_T, \eta ,\phi \right)_k \}
  \qquad \text{with} \qquad 
  k=1~...~n_C \; ,
\end{align}
so the jet phase space $\mathcal{J}$ has dimension $3 n_C$.  For the
training, we first sample a batch of jets $\{x_i\}$ from the dataset
and apply a set of symmetry-inspired augmentations to each jet to generate an
augmented batch $\{x_i^\prime\}$. Pairs of original and augmented
jets are defined as
\begin{alignat}{7}
  \text{positive pairs:} &\qqquad \{(x_i,x_i^\prime)\} \notag \\
  \text{negative pairs:} &\qqquad \{(x_i,x_j)\} \cup \{(x_i,x_j^\prime)\}
  \quad \text{for} \quad i \neq j \; .
\end{alignat}
The goal of the network training is to map positive pairs close
together in the representation space $\mathcal{R}$ and negative pairs
far apart.  This way, positive pairs are used to impose invariances
under symmetry transformations or theory augmentations of the jets in
$\mathcal{R}$, while the negative pairs are used to ensure that the
representation retains discriminative power within the dataset.  Truth
labels indicating if the jets are QCD or top are never used in the
optimization.

\subsubsection*{Loss function}

The mapping of Eq.\eqref{eq:mapping} defines the network outputs $z_i$
and $z_i^\prime$, each of them vectors describing jets in
$\mathbb{R}^{\text{dim}(z)}$. The actual representation, however, is
given by $f(x_i) = z_i/|z_i|$ and $f(x_i^\prime) =
z_i^\prime/|z_i^\prime|$, which means it is defined on a unit
hypersphere
\begin{align}
  \mathcal{R} = S^{\text{dim}(z)-1} \; .
\end{align}
On this sphere we define the similarity between two jets as~\cite{49372}
\begin{align}
s(z_i,z_j)=\frac{z_i\cdot z_j}{|z_i||z_j|}=\cos\theta_{ij} \; ,
\label{eq:cosine}
\end{align}
with $\theta_{ij}$ being the angle between the jets in $\mathcal{R}$.
The contrastive loss for a positive pair of jets is defined in terms
of this distance as
\begin{align}
\loss_{i} = -\log \frac{ e^{s(z_i,z_i^\prime)/\tau} }{ \sum_{j \neq i \in\text{batch}} \left[ e^{s(z_i,z_j)/\tau} + e^{s(z_i,z_j^\prime)/\tau} \right]} \; ,
\label{eq:clloss}
\end{align}
and the total loss is given by the sum over all positive pairs in the
batch, $\loss = \sum_i \loss_i$. Because the positive pairs appear in
the numerator, while the negative pairs contribute to the denominator,
the loss decreases when the distance between positive pairs becomes
smaller and when the distance between negative pairs becomes larger.
The hyper-parameter $\tau$ is referred to as the temperature and
controls the relative influence of positive pairs and
negative pairs.
The cosine similarity in Eq.\eqref{eq:cosine} is not a proper distance
metric, but we can define an angular distance as $d(z_i,z_j) =
\theta_{ij}/\pi = 0~...~1$, such that it satisfies the triangle
inequality.  

\begin{figure}[t]
  \centering
  \includegraphics[width=0.7\textwidth]{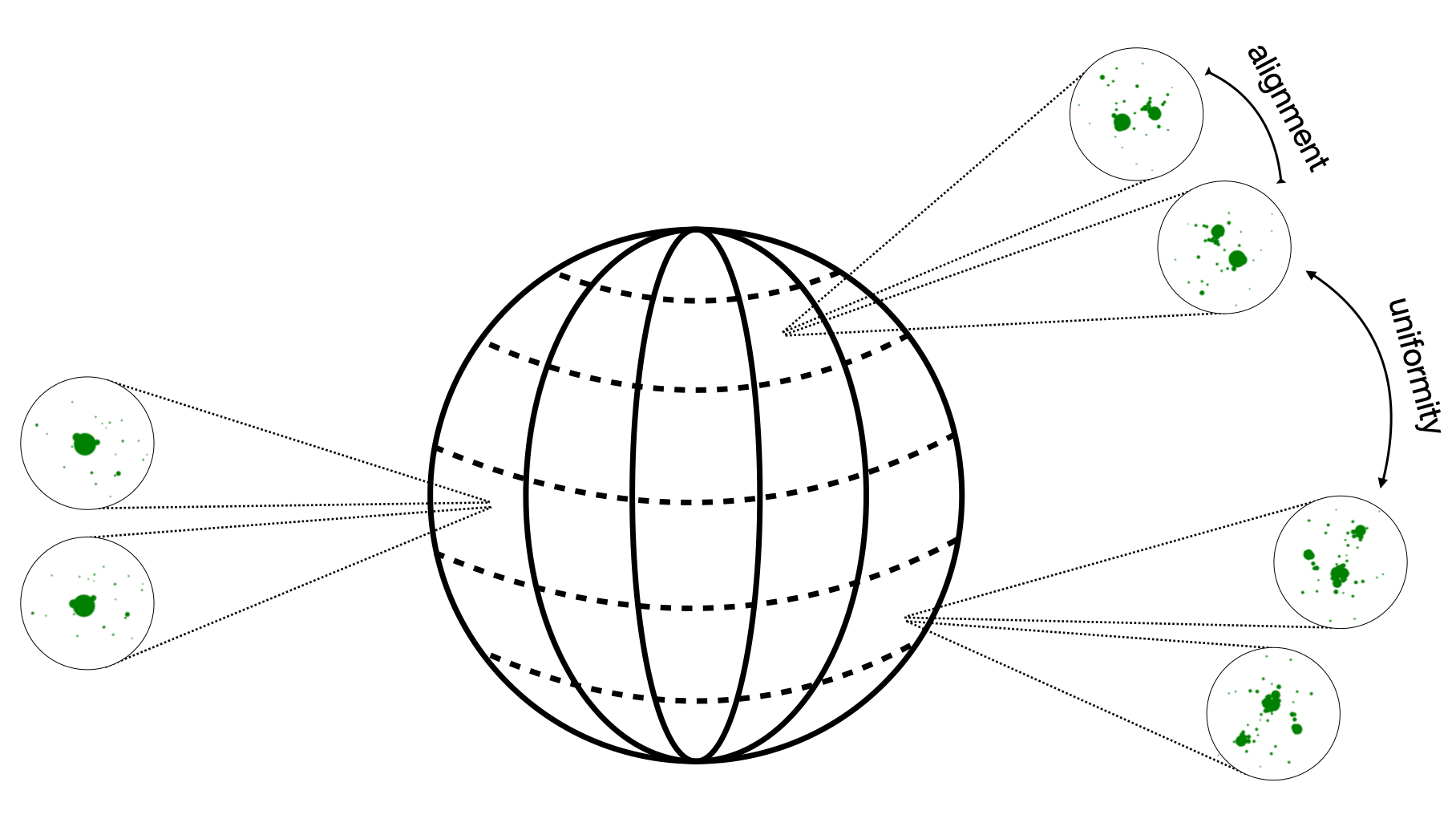}
  \caption{Illustration of the uniformity and alignment concepts
    behind the contrastive learning.}
  \label{fig:sphere}
\end{figure}
 
\subsubsection*{Uniformity vs alignment}

The contrastive loss can be understood in terms of uniformity versus
alignment on the unit hypersphere defining $\mathcal{R}$, illustrated
in Fig.~\ref{fig:sphere}.  The numerator of Eq.\eqref{eq:clloss},
describing the positive pairs, is minimal when all jets and their
augmented counterparts are mapped to the same point,
$s(z_i,z_i^\prime)=1$.  On a hypersphere, the negative pairs cannot be
pushed infinitely far apart, as would be possible in
$\mathbb{R}^{\text{dim}(z)}$, so the corresponding loss is minimal
when the jets are uniformly distributed on the hypersphere. We can
measure uniformity and alignment through
\begin{align}
\loss_\text{align} & = \frac{1}{N_\text{batch}} \sum_{i\in\text{batch}} s(z_i,z_i^\prime) \notag \\
\loss_\text{uniform} & = \frac{1}{N_\text{batch}} \sum_{i\in\text{batch}} \log \sum_{j\neq i} \left[ e^{-s(z_i,z_j)}+e^{-s(z_i,z_j^\prime)} \right] \;.
\end{align}
While the alignment function has the trivial solution where all jets
and all augmented jets are mapped to the same point, the uniformity
function does not have such a solution.  To map the jets to a uniform
distribution in a high-dimensional space, the mapping must learn
features of the jets to discriminate between them and map them to
different points. Uniformity alone is a sufficient optimization task
to obtain a representation with discriminative power.  The additional
alignment condition develops a mapping to $\mathcal{R}$, which focuses
on the invariance with respect to augmentations and symmetries.  The
combined contrastive learning will not find representations which
are perfectly aligned or perfectly uniform.

\subsubsection*{Symmetries and augmentations}

The mapping to the representation space $\mathcal{R}$ is optimized to
be approximately invariant to pre-defined symmetry transformations and
data augmentations.  Before applying symmetry transformations and
augmentations in the contrastive learning method we center the
jets such that the $p_T$-weighted centroid is at the origin in the
$\eta - \phi$ plane.

Rotations around the jet axes turn out to be a very efficient symmetry
we can impose on our representations.  In the jet image representation
this is included through pre-processing, where each jet is centered
and then rotated such that its principal axis points at 12 o'clock.
Energy flow polynomials are rotationally invariant by construction,
since they are built from angular distances between the jet
constituents. We apply rotations to a batch of jets by rotating each
jet through angles sampled from $0~...~2\pi$. Such rotations in the
$\eta - \phi$ plane are not Lorentz transformations and do not
preserve the jet mass, but for narrow jets with $R\lesssim1$ the
corrections to the jet mass can be neglected.

As a second symmetry we implement translations in the $\eta - \phi$
plane.  To do so, all constituents in a jet are shifted by the same
random distance, where shifts in each direction are limited to between
$-1~...~1$.  This performs better than restricting to smaller shifts.

In addition to (approximate) symmetries, we also employ
theory-inspired augmentations.  The distinction between the two is
much clearer in our physics application than it is in traditional
machine learning.  Quantum field theory tells us that soft gluon
radiation is universal and factorizes from the hard physics in the jet
splittings.  To encode this invariance in $\mathcal{R}$ we augment our
jets by smearing the positions of the soft constituents, \ie by
re-sampling the $\eta$ and $\phi$ coordinates of each constituent from
a Gaussian distribution centred on the original coordinates,
\begin{align}
  \eta^\prime \sim \mathcal{N} \left( \eta, \frac{ \Lambda_\text{soft}}{ p_T } \right)
  \qquad \text{and} \qquad
  \phi^\prime \sim \mathcal{N} \left( \phi, \frac{ \Lambda_\text{soft}}{ p_T } \right) \; ,
\end{align}
with a $p_T$-suppression in the variance relative to
$\Lambda_\text{soft}=100$~MeV.  

Similar to soft splittings, also collinear splitting lead to
divergences in perturbative quantum field theory. In practice. they
are removed through the finite angular resolution of a detector, which
will not be able to distinguish two constituents with $p_{T,a}$ and
$p_{T,b}$ at vanishing separation $\Delta R_{ab}\ll1$. We introduce
collinear augmentations to encode this feature by selecting
constituents and splitting them such that the total $p_T$ in an
infinitesimal region of the detector is unchanged,
\begin{align}
p_{T,a}+p_{T,b} = p_T \qqqquad
\eta_a = \eta_b &= \eta \notag \\
\phi_a = \phi_b &= \phi \; .
\end{align}
Our soft and collinear augmentations will enforce an approximate
IRC-safety in the jet representation. Unlike for instance EFPs, we do
not explicitly enforce it through a fixed set of angular correlations
or $p_T$-scalings, but let the contrastive optimization determine the
mapping to the representation space.

\section{JetCLR}
\label{sec:jetclr}

The symmetries discussed in the last section leave out one of the key
symmetries in jet representations, namely permutation symmetry. We
will include it through the transformer architecture, mapping jet
phase space to representation space. The combination of contrastive
loss and a permutation-invariant network architecture defines our
JetCLR concept.

\subsubsection*{Attention}

The key feature of transformer networks is
attention~\cite{bahdanau2014neural, luong2015effective}, more
specifically self-attention, which is an operation on a set of elements. 
Attention allows an element
of the set to assign weights to other elements. These
weights are then multiplied with some other piece of information from
the elements, so that more `attention' is placed on those elements
which achieve a high weight. We use the scaled dot-product
multi-headed self-attention of Ref.~\cite{vaswani2017attention}.

\begin{figure}[b!]
  \centering
  \includegraphics[width=0.7\textwidth]{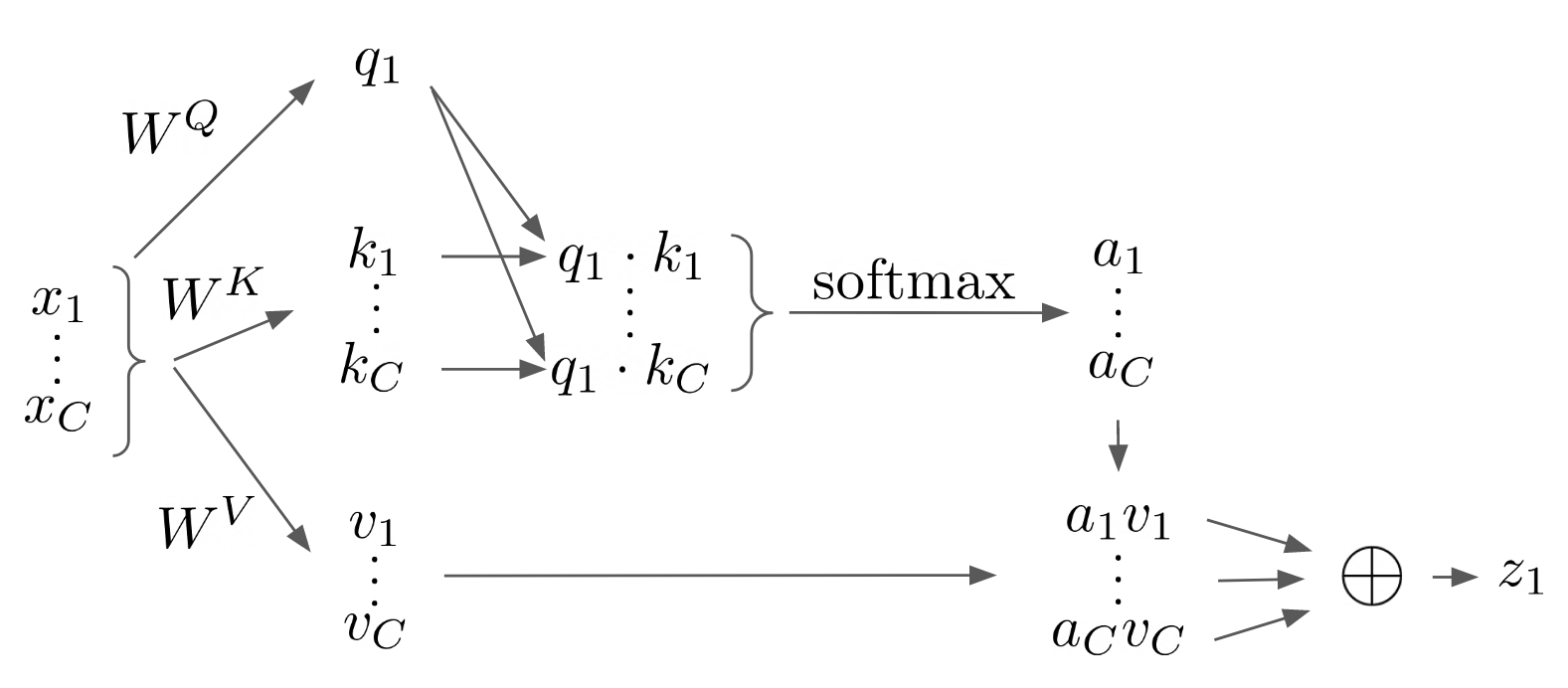}
  \caption{Illustration of single-headed self-attention. All elements
    are defined in the text.}
  \label{fig:attn}
\end{figure}

We illustrate the single-headed attention mechanism in
Fig.~\ref{fig:attn}. In this case, we apply it to a single jet
constituent described by the phase space coordinate $x_i$, where we
slightly abuse our notation such that the index now refers to
constituents rather than jets and in the following we will just
consider $x_1$ for one of the jets.  The learned weight matrix $W^Q$
transforms a single input $x_1$ to the corresponding query $q_1 = W^Q
x_1$. The complete set of inputs $x_1~...~x_C$ is transformed into
keys $k_1~...~k_C$ through a learned matrix $W^K$. Queries and keys
are then combined through a scalar product, which is scaled by the
dimension $d$ of $q_i$, and normalized by a softmax function to a set
of weights $a_i \in [0,1]$. Finally, the complete set of inputs
$x_1~...~x_C$ is transformed into a set of values $v_1~...~v_C$
through a third learned matrix $W^V$, and these $v_i$ are weighted by
the $a_i$ to provide a network output
\begin{align}
  z_1 = \sum_i \text{softmax}_i (q_1, k_i) \, v_i
  = \sum_i \text{softmax}_i \left( \frac{(W^Q x_1) \cdot (W^K x_i)}{\sqrt{d}} \right) \, W^V x_i
  \; . 
\end{align}
This can be thought of as a projection onto the basis $v_i$, where the
coefficients are given by the $\text{softmax}
(q_1, k_i)$ between $q_1$ and the $k_i$.  The equivalent operation is
applied to all $x_j$, leading to a set of outputs $z_j$.  Due to the
sum over set elements, each output $z_i$ is invariant to the
permutation of the other elements of the set, meaning that the entire
self-attention operation is permutation equivariant.

A problem with the self-attention mechanism is that each element of
the sequence tends to attend dominantly to
itself~\cite{vaswani2017attention}. This can be solved by extending
the network to multiple heads, where we perform several self-attention
operations in parallel, each with separate learned weight matrices,
then concatenate the outputs before applying a final linear layer.  In
practice, the full calculation for all constituents, all attention
heads, and for an entire batch is carried out in parallel with tensor
operations.

\subsubsection*{Transformer encoder}

In general, transformer networks include a complete encoder-decoder
architecture. In our application, we are only interested in deriving a
representation $\mathcal{J} \to \mathcal{R}$, so we use only the
encoder part of Ref.~\cite{vaswani2017attention}. It is a
sequence-to-sequence operation, made up of $N$ structurally identical,
successive blocks.

\begin{figure}[t]
  \centering
  \includegraphics[width=0.7\textwidth]{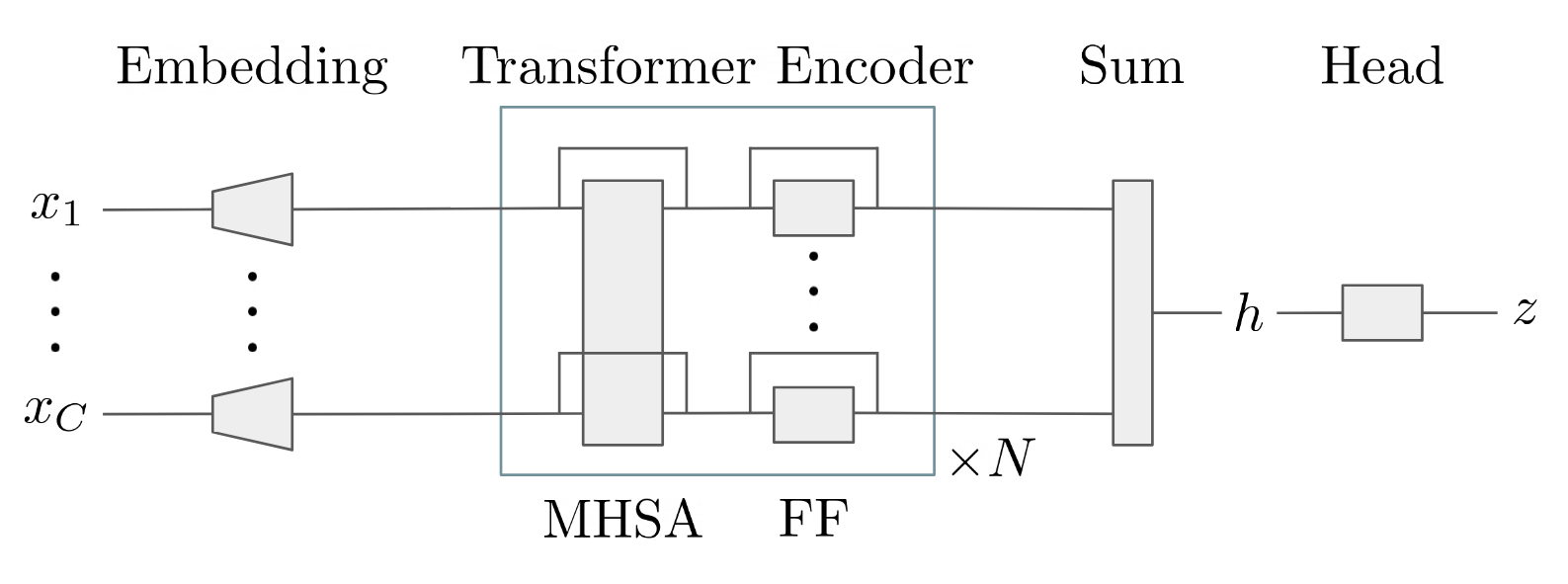}
  \caption{Illustration of the transformer network architecture. MHSA
    stands for multi-headed self-attention, and FF for a feed-forward
    block, as defined in the text.}
  \label{fig:net-diagram}
\end{figure}

The starting point is a set of constituents $x_i$, which we embed into
a higher-dimensional space by a single learned linear layer without
activation. This increases the representational power of the network.
A typical dimension of the embedding space is 1000.  Working with the
embedded jet constituents, each block contains the following
operations: multi-headed self-attention is applied to the input
constituent, and the result is added to the input, in a residual
fashion. This output is normalized using layer
normalization~\cite{ba2016layer} and passed through a residual
feed-forward network, which operates on each constituent
individually. The transformer--encoder block is repeated $N$ times.
Finally, the output is normalized using layer normalization.  The
encoder outputs are summed over constituents to produce a fixed-size
output $h$, which is passed through a final feed-forward head network
to give the output $z$. In practice, our supervised linear classifier
test will find that $h$ is a better representation than $z$,
consistent with typical practice in the self-supervised
literature~\cite{49372}.

Our network is implemented in PyTorch~\cite{paszke2019pytorch} with
the TransformerEncoder module, we also make heavy use of
NumPy~\cite{2020NumPy-Array}.

\subsubsection*{Variable-length inputs}

A general feature of jet constituents is that their number per jet is
variable. As in all ML tools for jet analyses, we zero-pad jets with
fewer constituents. This makes it easier to convert a batch to a
single tensor input for efficient computation and allows us to
concatenate the batch elements with equal length.  To ensure that this
padding does not affect the network output, we implement masking in
the transformer.  To stop information flow from zero-valued
constituents, we require the attention weights corresponding to those
constituents to be zero, technically by adding negative infinity to
the attention weight before the softmax normalization.  In addition,
we remove zero constituents from the sum over constituents to ensure
that the transformer is completely invariant to zero padding.

This masking ensures that constituents with zero $p_T$ have no effect
on the output, but we can generalize this by defining the masking to
be continuous in $p_T$. Instead of adding negative infinity to some
pre-softmax attention weights, we add $\beta \log p_T$ to all
pre-softmax attention weights. In addition, instead of setting some
transformer outputs before summation to zero, we multiply all
transformer outputs by the input $p_T$.  This IR-safe attention
mechanism renders the transformer network IR-safe by construction.

\section{Pretty good results}
\label{sec:results}

\begin{table}[t]
\centering
\begin{small}
\begin{minipage}[c][1.45in][t]{0.45\textwidth}
\begin{tabular}{l|cc}
\toprule
hyper-parameter 				& value	\\
 \midrule
  Model (embedding) dimension			&  1000	\\
  Feed-forward hidden dimension			&  same	\\
  Output dimension				&  same	\\
  \# self-attention heads			&  4	\\
  \# transformer layers	($N$)			&  4	\\
  \#

   layers				&  2	\\
  Dropout rate					&  0.1	\\
\bottomrule
\end{tabular}
\end{minipage}
\begin{minipage}[c][1.45in][t]{0.45\textwidth}
\begin{tabular}{l|cc}
\toprule
hyper-parameter 				& value	\\
 \midrule
  Optimizer					&  Adam($\beta_1 = 0.9, \beta_2 = 0.999$)\\
  learning rate					&  $5\times 10^{-5}$ 			\\
  batch size					&  128					\\
  \# epochs					&  500					\\
\bottomrule
\end{tabular}
\end{minipage}
\end{small}
\caption{Default setup of the transformer encoder and the training,
  unless noted explicitly.}
  \label{tab:arch}
\end{table}

After introducing all JetCLR elements, we have to investigate how its
various symmetries and augmentations contribute to its performance and
how our representation compares to alternative approaches.

Our transformer setup is given in Tab.~\ref{tab:arch}.  The
temperature hyper-parameter $\tau$ determines the trade-off between
the alignment and uniformity. It can strongly affect the performance
of the representations in a LCT.  We find that $\tau = 0.1~...~0.2$
works best which, despite the very different applications, is in
agreement with Refs~\cite{49372,wang2021understanding}.  We also see
that more model dimensions result in better performance, although with
the transformer network this performance gain seems to plateau around
$1000~...~1500$ dimensions.  Earlier tests using a fully connected
network instead of a transformer indicated that this plateau happens
at around 200 dimensions. We focus on the $1000$-dimensional
representations because this will eventually provide a fair comparison
to the EFPs at $d\leq7$, which is also a $1000$-dimensional
representation of the jets.

Our LCT is a linear neural network with a binary cross-entropy loss, optimized using stochastic
gradient descent. The network is trained with 50k top and QCD jets each for 5000
epochs with a batch size of 2056.  The exact set-up along with some alternative LCT setups are discussed in the
Appendix, with a focus on their respective strengths and underlying assumptions.

\begin{table}[t]
\centering
\begin{small}
\begin{minipage}[c][1.3in][t]{0.60\textwidth}
\begin{tabular}{l|cc}
\toprule
Augmentation 					& $\epsilon^{-1}(\epsilon_s\!=\!0.5)$  		& AUC					  		\\
 \midrule
 none 						& 15 								& 0.905 						\\
 translations					& 19 								& 0.916						\\
 rotations						& 21 								& 0.930						\\
 soft+collinear					& 89 								& 0.970						\\
 all combined (default)					& 181							& 0.979						\\
\bottomrule
\end{tabular}
\end{minipage}
\begin{minipage}[c][1.3in][t]{0.28\textwidth}
\begin{tabular}{c|cc}
\toprule
S/B 		& $\epsilon^{-1}(\epsilon_s\!=\!0.5)$  		& AUC  						\\
 \midrule
1.00 		& 181 							& 0.980 						\\
0.50		& 160 							& 0.979						\\
0.25		& 150							& 0.978						\\
0.10		& 161							& 0.978						\\
0.05		& 146							& 0.978						\\
0.01		& 158							& 0.978 \\
\bottomrule
\end{tabular} 
\end{minipage}
\end{small}
\caption{Left: classification results for JetCLR trained with
  different symmetries and augmentations and $S/B\!=\!1$. The default
  setup includes translation and rotation symmetries, combined with
  soft and collinear augmentations. Right: classification results for
  the combined (default) symmetries and augmentations, trained with
  different $S/B$.}
\label{tab:aug}
\end{table}

\subsubsection*{JetCLR}

From first principles, it is not clear which symmetries and
augmentations work best for learning representations with JetCLR.  In
the left panel of Tab.~\ref{tab:aug} we summarize the results after
applying rotational and translational symmetry transformations and
soft+collinear augmentations. To get an idea, we quote the best of a
number of runs for each case. Individually, the soft+collinear
augmentation works best.  Translations and rotations are less powerful
individually, but the combination of all three provides by far the
best representations.  The results for the individual augmentations in
Tab.~\ref{tab:aug} were obtained using regular masking in the
transformer. When combining all symmetries and augmentations the IR-safe
masking gives a slight boost, so our default in Tab.~\ref{tab:aug}
includes IR-safe attention.

While our initial results are based on a dataset containing equal
amounts of QCD and top jets, any application to anomaly detection
requires our approach to work with much fewer top signal jets. In the
right panel of Tab.~\ref{tab:aug} we show the performance of our
default benchmark for a decreasing fraction of signal events in the
training sample.  For each signal model with $S/B\!<\!1$ we only train one model, so we
expect some noise in the results.  The outcome indicates that the
JetCLR performance in the LCT is hardly sensitive to the amount of
signal jets in the training data, and that JetCLR can encode its
fundamental structures based on QCD jets and the symmetries and
augmentations alone.  Due to the stochastic nature of jet data, no
pattern is exclusive to top jets, so the QCD jet sample should indeed
contain all relevant information.  This result is very promising for
future anomaly searches using JetCLR representations.

\begin{figure}[t]
  \includegraphics[width=0.45\textwidth]{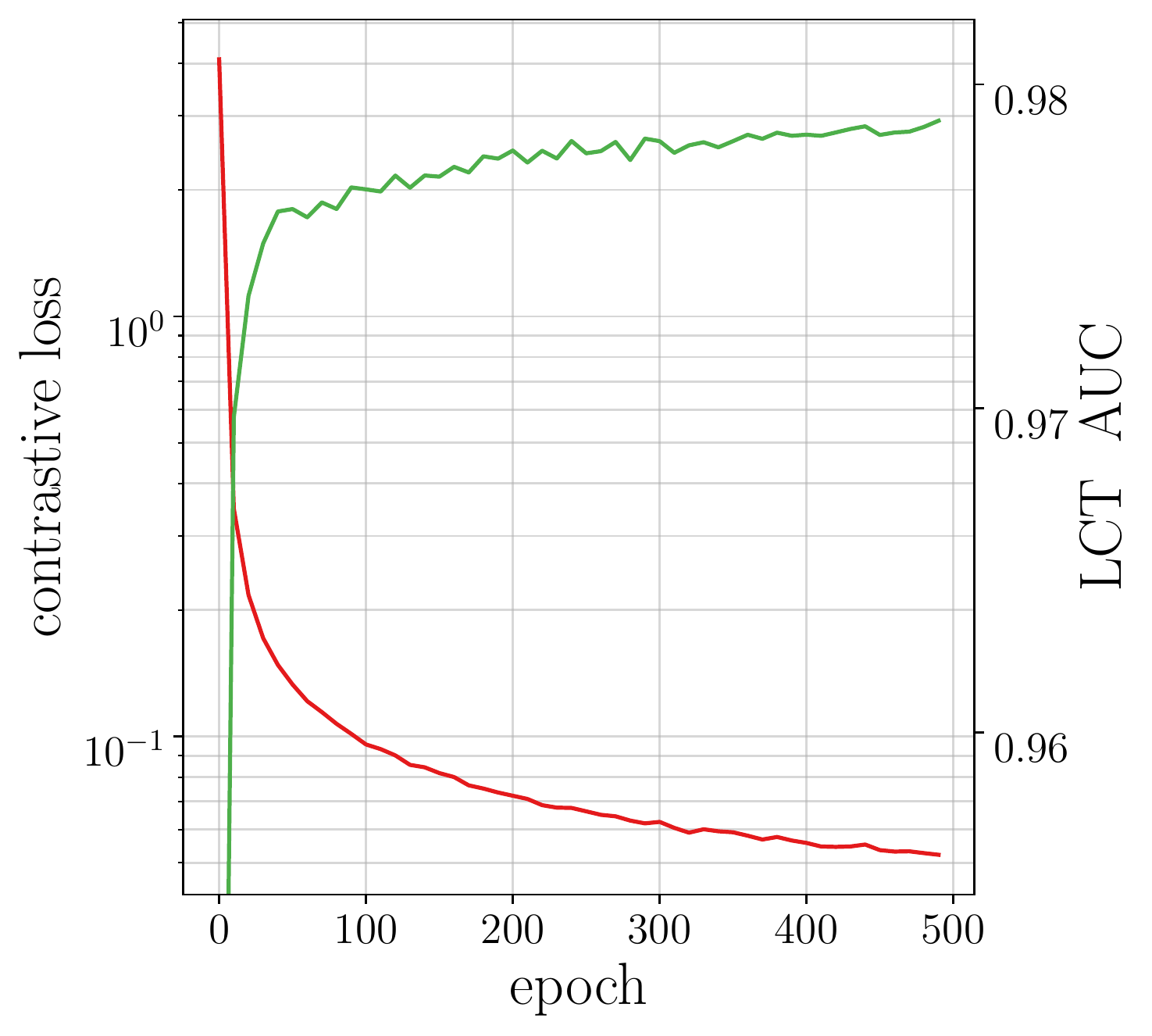}
  \hspace*{0.05\textwidth}
  \includegraphics[width=0.45\textwidth]{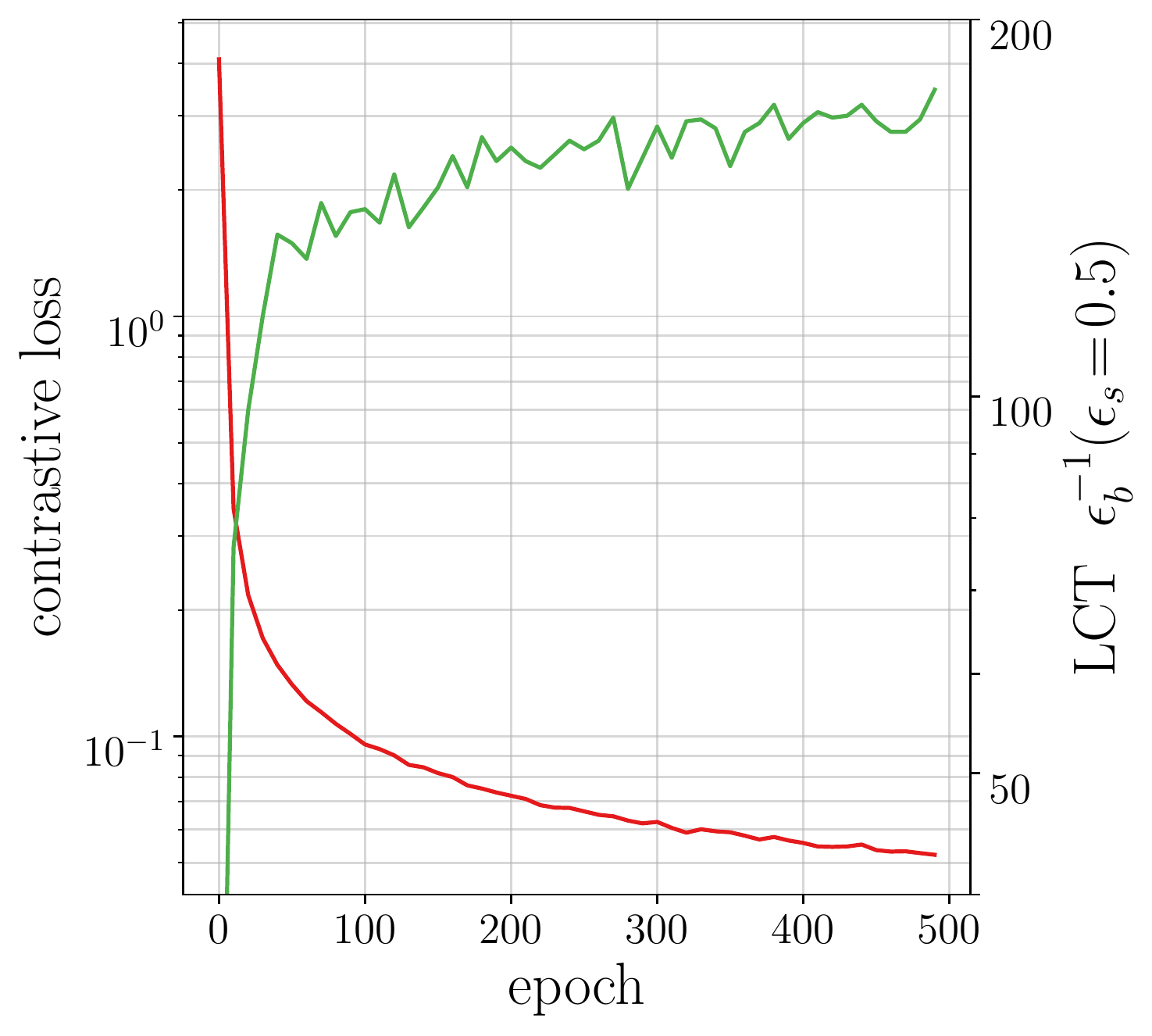}
  \caption{Loss and LCT performance on test data as a function of the
    training epochs.  The LCT is performed every $10$ epochs.}
\label{fig:loss-auc}
\end{figure}

To test our JetCLR training we analyse the AUC and the mistagging rate
of the LCT on the test data as a function of the training epoch. Given
the impressive LCT scores we could just assume that optimizing the
contrastive loss is a good auxiliary task for constructing good
representations for classification. However, for anomaly detection we
know that the auxiliary optimization task may appear to be converging
to a good representation for classification initially, but then
diverge at larger epochs. In Fig.~\ref{fig:loss-auc} we show that the
increasing performance of the JetCLR representations in the LCT is
indeed aligned with the optimization of the contrastive loss function.

\subsubsection*{Encoded symmetries}

\begin{figure}[t]
  \centering
  \includegraphics[width=0.40\textwidth]{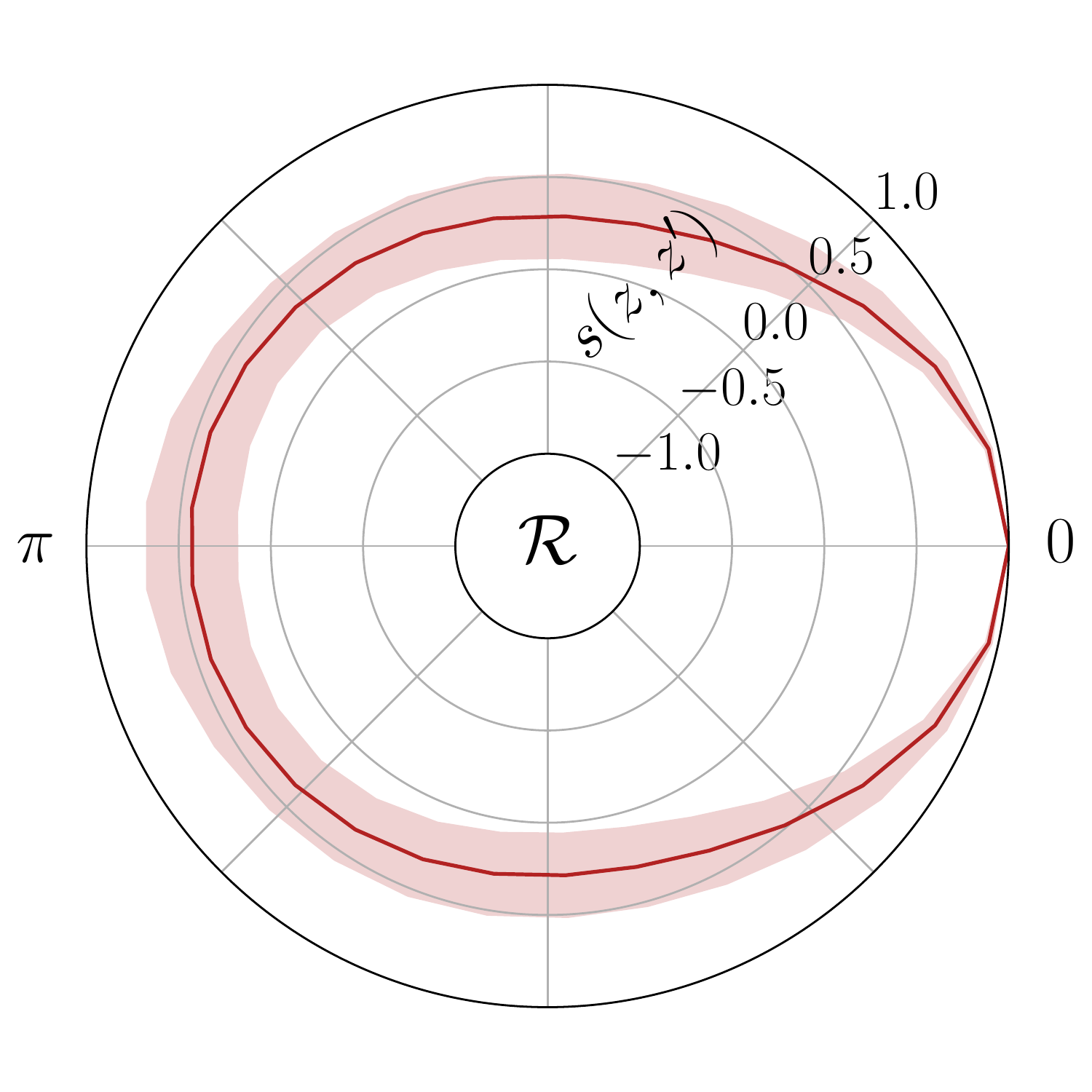}
  \hspace*{0.05\textwidth}
  \includegraphics[width=0.40\textwidth]{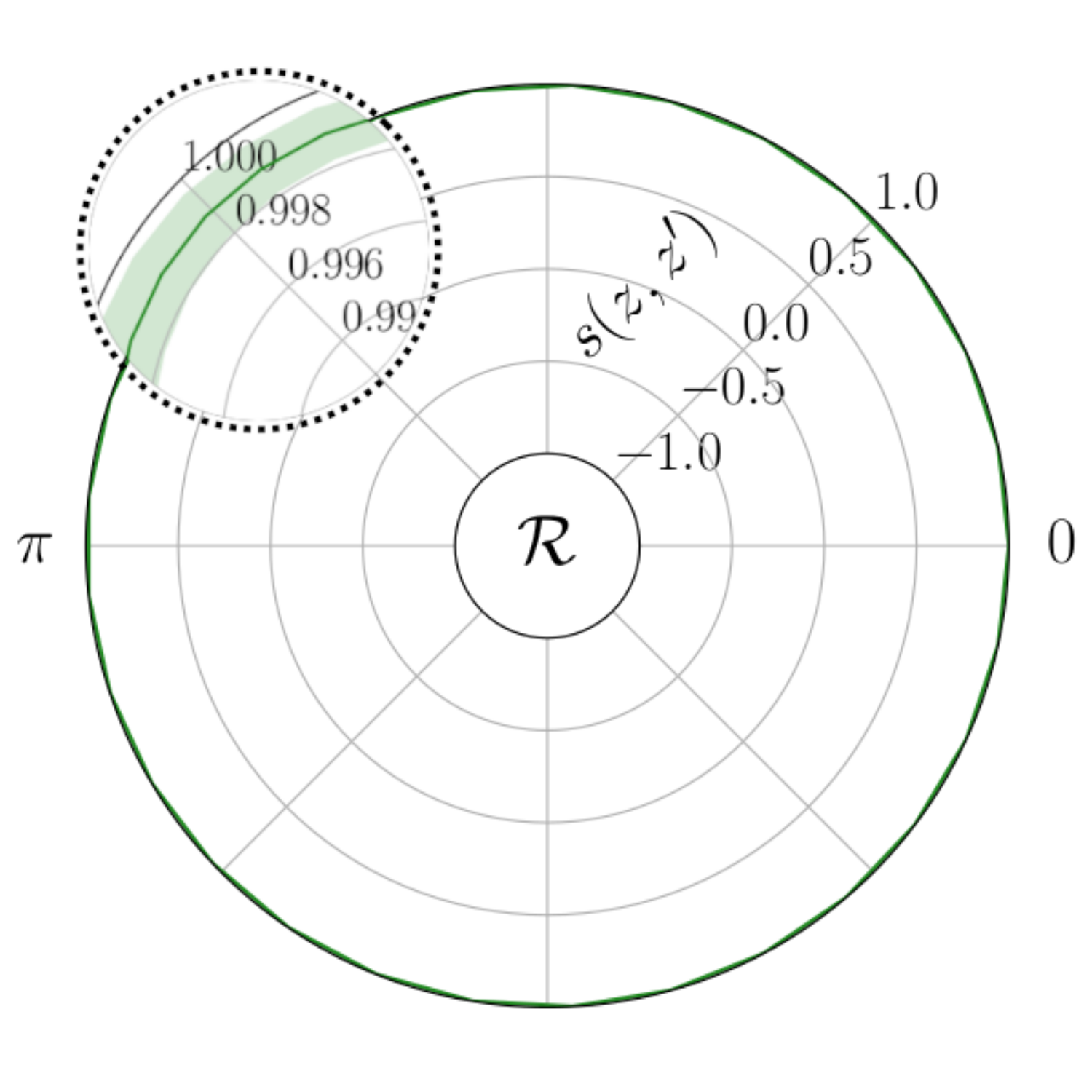}
  \includegraphics[width=0.40\textwidth]{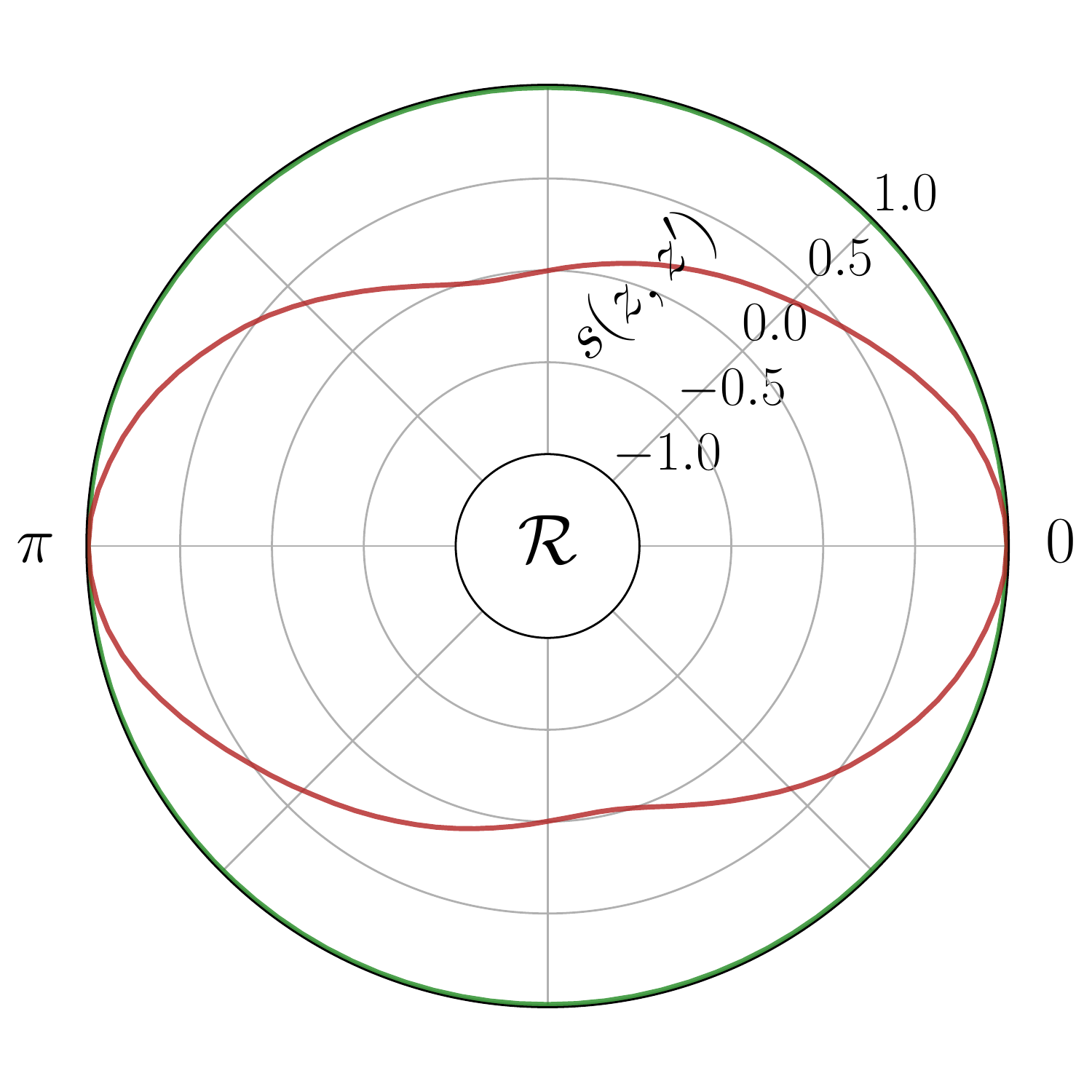}
  \hspace*{0.05\textwidth}
  \includegraphics[width=0.40\textwidth]{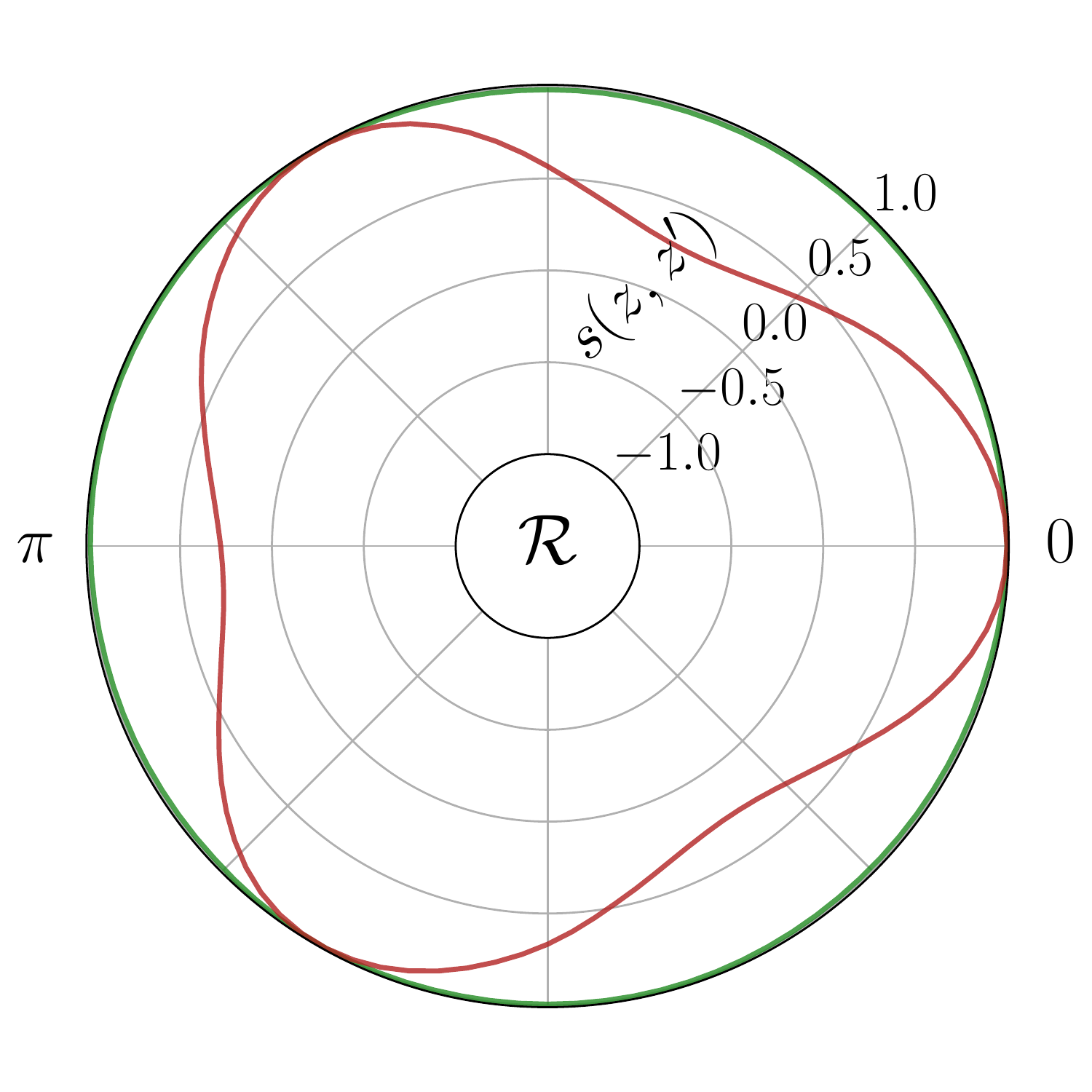}
  \caption{Visualization of the rotational invariance in
    representation space, keeping in mind that $s(z,z') = 1$ indicates
    identical representations.  Top: JetCLR representation trained
    without (left) and with (right) rotational transformations. Note
    the different scales of the radial axes showing the jet similarity
    defined in Eq.\eqref{eq:cosine}.  Bottom: JetCLR representation
    for 2-prong (left) and 3-prong (right) toy jets, trained without
    (red) and with (green) rotational transformations.}
  \label{fig:rot-radial}
\end{figure}

Of the two basic JetCLR tasks, invariance and discriminative power, we
first confirm that the network indeed encodes symmetries.  To
illustrate the encoded rotation symmetry we show how the
representation is invariant to actual rotations of jets.  We start
with a batch of 100 jets, and produce a set of rotated copies for each
jet, with rotation angles evenly spaced in $0~...~2\pi$.  We then pass
each jet and its rotated copy through the network, and calculate their
cosine similarity, Eq.\eqref{eq:cosine}, with the original jet.  In
the top panels of Fig.~\ref{fig:rot-radial} we show the mean and
standard deviation of the cosine similarity as a function of the
rotation angle. First, from the scale of the radial axis $s(z,z')$ we
see that the representations obtained by training JetCLR with
rotations are much more similar to the original jets. Second, in the
left panel the similarity varies between $0.5$ and $1.0$ as a function
of the rotation angle, while in the right panel the JetCLR
representation is indeed rotationally invariant.

Next, we create toy jets with $p_{T,j}=600$~GeV, one with two
constituents and one with three equally spaced constituents. The jet
momentum is shared equally between the subjets. We then compare how
rotationally invariant their JetCLR representations are in the lower
panels of Fig.~\ref{fig:rot-radial}. The red lines represent the
similarity functions for JetCLR representations of two-prong (left)
and three-prong (right) jets, trained without rotational
transformations. The maximum values of $s(z,z')$ reflect the
degeneracies from the geometric symmetry of the toy jets.  The green
line represents the similarity function for the JetCLR representations
trained with rotational transformations.

\subsubsection*{JetCLR performance}

\begin{figure}[t]
  \centering
  \includegraphics[width=0.60\textwidth]{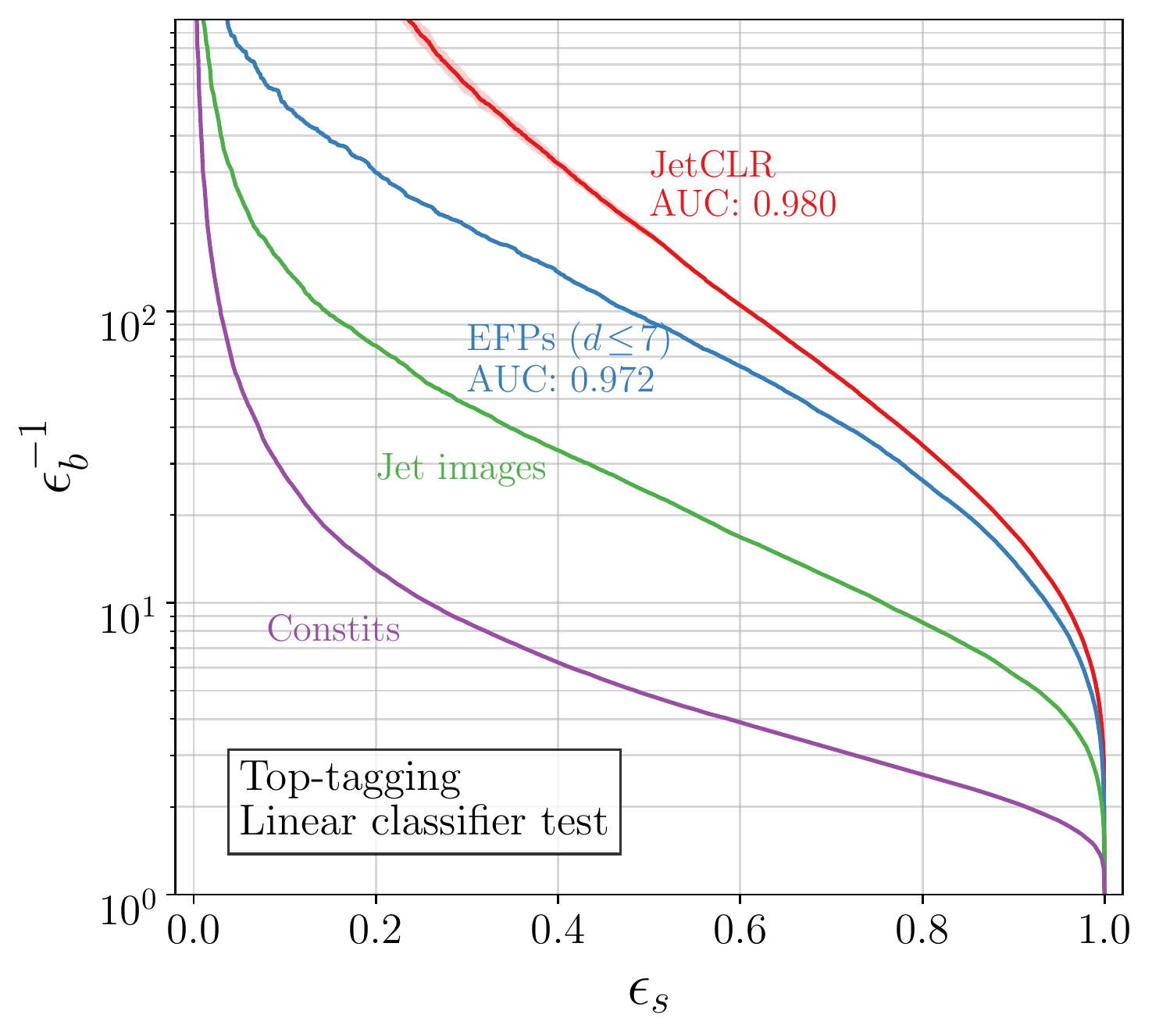}
  \caption{Comparison of JetCLR with other classification metrics.}
  \label{fig:roc_comp}
\end{figure}

After confirming that the JetCLR indeed encodes symmetries, we turn to
the second task, namely discriminative power.  To put the results of
Tab.~\ref{tab:aug} into context, we show ROC curves for JetCLR and
various other representations in Fig.~\ref{fig:roc_comp}. For the
constituents representation we take the 20 hardest constituents in
each jet, flatten their $(p_T,\eta,\phi)$ components into a single
vector, and feed them to the linear classifier. For the jet images
representation we use the pre-processing of
Ref.~\cite{Heimel:2018mkt,Farina:2018fyg}, flattening the $40\times
40$ image to a single 1600-dimensional vector and giving it to the
linear classifier. Finally, the EFP representation is invariant to
permutations and to rotations by construction, and its IR-safety
guarantees independence from soft activity. In many ways, EFPs can be
considered a theory-driven counterpart of our JetCLR tool.  We use all
coefficients up to degree seven, since it has been shown that adding
higher powers does not improve the top-tagging~\cite{Komiske:2017aww},
and choose $\beta = 0.5$ for the exponent of the $p_T$-weights.  For
the JetCLR representation we used the default setup, restricting the
maximum number of constituents per jet to 50, and masking jets with
fewer constituents.

For all representations we train the networks with $100$k jets split
evenly between top and QCD.  For the alternative representations we
run eight linear classifiers and use the mean over the mistag rates
for the ROC curve.  For the JetCLR representation an additional source
of uncertainty arises from the training of the transformer-encoder
network. We train two linear classifiers on four different
representations from four different JetCLR runs and show the mean and
standard deviation of the mistag rate vs the efficiency.

As expected, representations using more knowledge of the physical
symmetries perform increasingly well.  The top-performing EFP and
JetCLR representations use the same latent dimension, and the
self-supervised JetCLR method slightly outperforms the systematic EFP
representation for a linear network with a binary cross-entropy loss,
the LCT with the weakest assumptions about the data.  As discussed in
the Appendix, the EFP results improve for a linear discriminant
analysis, where the underlying assumption on the data is not
applicable to JetCLR.  Obviously, any LCT can only serve as a proxy to
estimate representation quality, the real test will be performance in
an actual analysis.

\section{Conclusions}
\label{sec:cln}

We have introduced Contrastive Learning (CLR) to design observables
which respect symmetries and data augmentations while retaining
discrimination power within the dataset.  We have applied this new
method in jet physics, developing the JetCLR tool\footnote{The JetCLR code will be maintained at
  \href{https://github.com/bmdillon/JetCLR}{https://github.com/bmdillon/JetCLR}}.
Guided by fundamental symmetries and principles of quantum field
theory, we introduced a transformer-encoder network to encode
rotation, translation, and permutation symmetries, as well as
invariance under soft and collinear constituent augmentations.

After visualizing the symmetry-enhanced representation space, we
evaluated the network performance using a linear classifier test, a
simple supervised classifier trained on the representations to
distinguish top jets from QCD jets.  Due to the simplicity of the
classifier, its performance can be interpreted as a quality measure
for the representations. We find that self-supervised JetCLR
outperforms simple jet images and is competitive with energy flow
polynomials.

Regardless of our specific JetCLR application, our key point is that
it is possible to incorporate symmetry principles and physics
knowledge in self-supervised ML-tools and latent representations. This
opens many avenues for future work with JetCLR and contrastive
learning in general. Because of the way JetCLR incorporates symmetries
from a single augmented data set, it is particularly well suited to
enhance and control anomaly searches, one of the great
ML-opportunities for future LHC runs.

\begin{center} \textbf{Acknowledgments} \end{center}

We are, again, grateful to Ullrich K\"othe for invaluable discussions.
The research of TP is supported by the Deutsche Forschungsgemeinschaft
under grant 396021762 -- TRR~257 \textsl{Particle Physics
  Phenomenology after the Higgs Discovery}. PS is partly supported by
the DFG Research Training Group GK-1940, \textsl{Particle Physics
  Beyond the Standard Model}.  BMD is supported by funding from BMBF.
GK acknowledges the support of the Deutsche Forschungsgemeinschaft
under Germany’s Excellence Strategy – EXC 2121 \textsl{Quantum
  Universe} – 390833306.

\appendix
\section{Linear classifier tests}

Without a direct application to a specific task, comparing data
representations is difficult.  Downstream tasks can vary from anomaly
detection to classification, or regression.  One standard method for
comparing representations is the linear classifier test (LCT), but
even this test can be ambiguous.  The idea behind the LCT is that the
linearity of the classifier removes much of the expressive power from
the classifier, so a linear classifier measures the expressive power
of the representation.  However, we find that in removing expressive
power from the classifier, the results become much more dependent on
the inductive biases incurred in the choice of loss function and
optimization. We discuss a few different LCTs and explain the
assumptions they make about the data they are optimized on. We provide
a more complete comparison between the JetCLR and EFP representations
using different LCTs in Tab.~\ref{tab:lcts}. All classifiers are
trained using 10-fold cross validation to identify the best-performing
hyperparameters. The reported performance is the average over the 10
folds.

\subsubsection*{Binary cross entropy loss}

A linear classifier trained with binary cross-entropy loss, also known
as logistic regression, makes an assumption about how the probability
of each of the two classes changes in different parts of the space. If
$x$ denotes data and $y \in \{0, 1\}$ the two classes, the assumption
is that
\begin{align}
  p(y=1 | x)
  = \text{sigmoid} (w^T x + c)
  = \frac{1}{1 + e^{-w^Tx - c}} \; ,
\end{align}
where $w$ is some vector and $c$ is a scalar bias. We find $w$ and $c$
by minimizing the KL-divergence between this model and the labeled
data. In practice, this means minimizing the binary cross entropy
\begin{align}
  \loss = \XLangle -\log \text{sigmoid}(y(w^T x + c)) \XRangle_{x,y} + \lambda \lVert w \rVert^2,
\end{align}                         
where $\lambda \ge 0$ is a regularization parameter. Regularization is
not strictly necessary but can improve performance. It can be turned
off by setting $\lambda = 0$. We select the best $\lambda \in
\{10^{-6}, 10^{-4}, 10^{-2}\}$ by 10-fold cross validation.  This
optimization problem is convex, so it should always give the same
optimal $w$ and $c$ when using a standard algorithm such as gradient
descent.

\begin{table}[t]
\centering
\renewcommand{\arraystretch}{1}
\renewcommand{\tabcolsep}{3mm}
\begin{tabular}{l|cc|cc}
\toprule
& \multicolumn{2}{c}{EFPs ($d\!\leq\!7$)} & \multicolumn{2}{c}{JetCLR} \\
& $\epsilon^{-1}(\epsilon_s\!=\!0.5)$  & AUC & $\epsilon^{-1}(\epsilon_s\!=\!0.5)$ & AUC \\
 \midrule
Binary cross-entropy (Fig.~\ref{fig:roc_comp})		& 93	& 0.972		& 181	& 0.980		\\
SVM (hinge loss)					& 88 	& 0.971 	& 130	& 0.977		\\
SVM (squared hinge loss)			& 100 	& 0.971		& 169	& 0.979		\\
Linear discriminant analysis						& 165 	& 0.979		& 133	& 0.977		\\
\bottomrule
\end{tabular}
\caption{Comparison of JetCLR and energy flow polynomials as in
  Fig.~\ref{fig:roc_comp}, including different linear classifier
  tests.}
\label{tab:lcts}
\end{table}

\subsubsection*{Support vector machine}

A (linear) support vector machine (SVM) separates two classes with as
wide a margin as possible, aiming for robustness. When the two classes
are not linearly separable, as in our application, the SVM minimizes
how far on the wrong side of the decision boundary misclassified
points are, but it does not consider points which are safely on the
correct side of the boundary. This is in contrast to logistic
regression, which pushes points to the correct side of the decision
boundary no matter how far over it they already are. SVMs are
expressible as a convex problem and make no assumptions about the
distribution of the data. However, they involve tuning a
hyperparameter which determines how strictly misclassifications are
enforced. This will lead to different results depending on the choice
of hyperparameter, which must be selected based on performance on some
other metric, for instance a classification accuracy.  We consider two
variants of SVM, starting with the standard hinge loss
\begin{align}
	\loss = \XLangle \max(0, 1 - y(w^T x + c)) \XRangle_{x,y} + \lambda \lVert w \rVert^2 \; ,
\end{align}
where $\lambda > 0$ is the regularization parameter. The variant with
the squared hinge loss minimizes
\begin{align}
\loss = \XLangle \max(0, 1 - y(w^T x + c))^2 \XRangle_{x,y} + \lambda \lVert w \rVert^2 \; .
\end{align}
The two differ in how strongly they penalize distance from the
decision boundary.  The squared variant enacts a weaker penalty for
points which are only just on the correct side of the boundary, but a
stronger penalty for points which are on the incorrect side. We select
the best $\lambda \in \{10^{-6}, 10^{-4}, 10^{-2}\}$ by 10-fold cross
validation.

\subsubsection*{Linear discriminant analysis}

Linear discriminant analysis makes a stronger assumption, namely
Gaussian distributed data. The data is modeled as a Gaussian mixture
model with two equally likely classes, where the covariance matrix of
the two classes is the same.  The means $\mu_{0,1}$ and the covariance
$\Sigma$ are typically estimated from labeled data.  The Bayes-optimal
classifier is the log ratio of the two probabilities.  Gaussian log
probabilities are quadratic in $x$, so if the two covariance-terms
cancel we obtain a linear equation in $x$.  As with the previous
classifiers, the classification score can be written as $w^T x + c$,
where here $w = (\mu_1 - \mu_0)^T \Sigma^{-1}$ and $c = -w^T (\mu_0 +
\mu_1)/2$. No cross validation is necessary to select hyperparameters,
but the reported results are averages over 10-fold cross validation.
The basic assumption of Gaussian data is not fulfilled by the JetCLR
representations, which are optimized for uniformity on a unit
hypersphere, so the linear discriminant analysis will not capture the
structure of the JetCLR representation.

\bibliography{literature}

\providecommand{\href}[2]{#2}\begingroup\raggedright\begin{thebibliography}{10}

\bibitem{Noether1918}
E.~Noether, {\it Invariante variationsprobleme},
  \href{http://eudml.org/doc/59024}{Nachrichten von der Gesellschaft der
  Wissenschaften zu Göttingen, Mathematisch-Physikalische Klasse {\bfseries
  1918} (1918)  235}.

\bibitem{Krippendorf:2020gny}
S.~Krippendorf and M.~Syvaeri, {\it {Detecting Symmetries with Neural
  Networks}},  \href{http://arxiv.org/abs/2003.13679}{{arXiv:2003.13679
  [physics.comp-ph]}}.

\bibitem{Barenboim:2021vzh}
G.~Barenboim, J.~Hirn, and V.~Sanz, {\it {Symmetry meets AI}},
  \href{http://arxiv.org/abs/2103.06115}{{arXiv:2103.06115 [cs.LG]}}.

\bibitem{Maiti:2021fpy}
A.~Maiti, K.~Stoner, and J.~Halverson, {\it {Symmetry-via-Duality: Invariant
  Neural Network Densities from Parameter-Space Correlators}},
  \href{http://arxiv.org/abs/2106.00694}{{arXiv:2106.00694 [cs.LG]}}.

\bibitem{Krippendorf:2021uxu}
S.~Krippendorf, R.~Kroepsch, and M.~Syvaeri, {\it {Revealing systematics in
  phenomenologically viable flux vacua with reinforcement learning}},
  \href{http://arxiv.org/abs/2107.04039}{{arXiv:2107.04039 [hep-th]}}.

\bibitem{Heimel:2018mkt}
T.~Heimel, G.~Kasieczka, T.~Plehn, and J.~M. Thompson, {\it {QCD or What?}},
  \href{http://dx.doi.org/10.21468/SciPostPhys.6.3.030}{SciPost Phys.
  {\bfseries 6} (2019) 3, 030},
\href{http://arxiv.org/abs/1808.08979}{{arXiv:1808.08979 [hep-ph]}}.

\bibitem{Farina:2018fyg}
M.~Farina, Y.~Nakai, and D.~Shih, {\it {Searching for New Physics with Deep
  Autoencoders}},  \href{http://dx.doi.org/10.1103/PhysRevD.101.075021}{Phys.
  Rev. D {\bfseries 101} (2020) 7, 075021},
  \href{http://arxiv.org/abs/1808.08992}{{arXiv:1808.08992 [hep-ph]}}.

\bibitem{Nachman:2020lpy}
B.~Nachman and D.~Shih, {\it {Anomaly Detection with Density Estimation}},
  \href{http://dx.doi.org/10.1103/PhysRevD.101.075042}{Phys. Rev. D {\bfseries
  101} (2020)  075042},
  \href{http://arxiv.org/abs/2001.04990}{{arXiv:2001.04990 [hep-ph]}}.

\bibitem{Bortolato:2021zic}
B.~Bortolato, B.~M. Dillon, J.~F. Kamenik, and A.~Smolkovi\v{c}, {\it {Bump
  Hunting in Latent Space}},
  \href{http://arxiv.org/abs/2103.06595}{{arXiv:2103.06595 [hep-ph]}}.

\bibitem{Dillon:2021nxw}
B.~M. Dillon, T.~Plehn, C.~Sauer, and P.~Sorrenson, {\it {Better Latent Spaces
  for Better Autoencoders}},
  \href{http://arxiv.org/abs/2104.08291}{{arXiv:2104.08291 [hep-ph]}}.

\bibitem{Metodiev:2017vrx}
E.~M. Metodiev, B.~Nachman, and J.~Thaler, {\it {Classification without labels:
  Learning from mixed samples in high energy physics}},
  \href{http://dx.doi.org/10.1007/JHEP10(2017)174}{JHEP {\bfseries 10} (2017)
  174}, \href{http://arxiv.org/abs/1708.02949}{{arXiv:1708.02949 [hep-ph]}}.

\bibitem{Kasieczka:2020pil}
G.~Kasieczka, B.~Nachman, M.~D. Schwartz, and D.~Shih, {\it {Automating the
  ABCD method with machine learning}},
  \href{http://dx.doi.org/10.1103/PhysRevD.103.035021}{Phys. Rev. D {\bfseries
  103} (2021) 3, 035021},
  \href{http://arxiv.org/abs/2007.14400}{{arXiv:2007.14400 [hep-ph]}}.

\bibitem{Barron:2021btf}
J.~Barron, D.~Curtin, G.~Kasieczka, T.~Plehn, and A.~Spourdalakis, {\it
  {Unsupervised Hadronic SUEP at the LHC}},
  \href{http://arxiv.org/abs/2107.12379}{{arXiv:2107.12379 [hep-ph]}}.

\bibitem{Cogan:2014oua}
J.~Cogan, M.~Kagan, E.~Strauss, and A.~Schwarztman, {\it {Jet-Images: Computer
  Vision Inspired Techniques for Jet Tagging}},
  \href{http://dx.doi.org/10.1007/JHEP02(2015)118}{JHEP {\bfseries 02} (2015)
  118}, \href{http://arxiv.org/abs/1407.5675}{{arXiv:1407.5675 [hep-ph]}}.

\bibitem{deOliveira:2015xxd}
L.~de~Oliveira, M.~Kagan, L.~Mackey, B.~Nachman, and A.~Schwartzman, {\it
  {Jet-images — deep learning edition}},
  \href{http://dx.doi.org/10.1007/JHEP07(2016)069}{JHEP {\bfseries 07} (2016)
  069}, \href{http://arxiv.org/abs/1511.05190}{{arXiv:1511.05190 [hep-ph]}}.

\bibitem{Kasieczka:2017nvn}
G.~Kasieczka, T.~Plehn, M.~Russell, and T.~Schell, {\it {Deep-learning Top
  Taggers or The End of QCD?}},
  \href{http://dx.doi.org/10.1007/JHEP05(2017)006}{JHEP {\bfseries 05} (2017)
  006}, \href{http://arxiv.org/abs/1701.08784}{{arXiv:1701.08784 [hep-ph]}}.

\bibitem{Lin:2018cin}
J.~Lin, M.~Freytsis, I.~Moult, and B.~Nachman, {\it {Boosting $H\to b\bar b$
  with Machine Learning}},
  \href{http://dx.doi.org/10.1007/JHEP10(2018)101}{JHEP {\bfseries 10} (2018)
  101}, \href{http://arxiv.org/abs/1807.10768}{{arXiv:1807.10768 [hep-ph]}}.

\bibitem{Komiske:2016rsd}
P.~T. Komiske, E.~M. Metodiev, and M.~D. Schwartz, {\it {Deep learning in
  color: towards automated quark/gluon jet discrimination}},
  \href{http://dx.doi.org/10.1007/JHEP01(2017)110}{JHEP {\bfseries 01} (2017)
  110}, \href{http://arxiv.org/abs/1612.01551}{{arXiv:1612.01551 [hep-ph]}}.

\bibitem{Macaluso:2018tck}
S.~Macaluso and D.~Shih, {\it {Pulling Out All the Tops with Computer Vision
  and Deep Learning}},  \href{http://dx.doi.org/10.1007/JHEP10(2018)121}{JHEP
  {\bfseries 10} (2018)  121},
  \href{http://arxiv.org/abs/1803.00107}{{arXiv:1803.00107 [hep-ph]}}.

\bibitem{Henrion:DLPS2017}
I.~Henrion, K.~Cranmer, J.~Bruna, K.~Cho, J.~Brehmer, G.~Louppe, and
  G.~Rochette, {\it {Neural Message Passing for Jet Physics}},  {Proceedings of
  the Deep Learning for Physical Sciences Workshop at NIPS (2017)} (2017)  .
\newblock
\newblock
  \href{{https://dl4physicalsciences.github.io/files/nips_dlps_2017_29.pdf}}{2017}.

\bibitem{Qasim:2019otl}
S.~R. Qasim, J.~Kieseler, Y.~Iiyama, and M.~Pierini, {\it {Learning
  representations of irregular particle-detector geometry with
  distance-weighted graph networks}},
  \href{http://dx.doi.org/10.1140/epjc/s10052-019-7113-9}{Eur. Phys. J. C
  {\bfseries 79} (2019) 7, 608},
  \href{http://arxiv.org/abs/1902.07987}{{arXiv:1902.07987 [physics.data-an]}}.

\bibitem{Chakraborty:2019imr}
A.~Chakraborty, S.~H. Lim, and M.~M. Nojiri, {\it {Interpretable deep learning
  for two-prong jet classification with jet spectra}},
  \href{http://dx.doi.org/10.1007/JHEP07(2019)135}{JHEP {\bfseries 07} (2019)
  135}, \href{http://arxiv.org/abs/1904.02092}{{arXiv:1904.02092 [hep-ph]}}.

\bibitem{1808887}
J.~Shlomi, P.~Battaglia, and J.-R. Vlimant, {\it {Graph Neural Networks in
  Particle Physics}},  \href{http://arxiv.org/abs/2007.13681}{{arXiv:2007.13681
  [hep-ex]}}.

\bibitem{Louppe:2017ipp}
G.~Louppe, K.~Cho, C.~Becot, and K.~Cranmer, {\it {QCD-Aware Recursive Neural
  Networks for Jet Physics}},
  \href{http://arxiv.org/abs/1702.00748}{{arXiv:1702.00748 [hep-ph]}}.

\bibitem{Andreassen:2018apy}
A.~Andreassen, I.~Feige, C.~Frye, and M.~D. Schwartz, {\it {JUNIPR: a Framework
  for Unsupervised Machine Learning in Particle Physics}},
\href{http://arxiv.org/abs/1804.09720}{{arXiv:1804.09720 [hep-ph]}}.

\bibitem{Dillon:2019cqt}
B.~M. Dillon, D.~A. Faroughy, and J.~F. Kamenik, {\it {Uncovering latent jet
  substructure}},  \href{http://dx.doi.org/10.1103/PhysRevD.100.056002}{Phys.
  Rev. {\bfseries D100} (2019) 5, 056002},
\href{http://arxiv.org/abs/1904.04200}{{arXiv:1904.04200 [hep-ph]}}.

\bibitem{Dillon:2020quc}
B.~M. Dillon, D.~A. Faroughy, J.~F. Kamenik, and M.~Szewc, {\it {Learning the
  latent structure of collider events}},
  \href{http://dx.doi.org/10.1007/JHEP10(2020)206}{JHEP {\bfseries 10} (2020)
  206}, \href{http://arxiv.org/abs/2005.12319}{{arXiv:2005.12319 [hep-ph]}}.

\bibitem{Dreyer_2018}
F.~A. Dreyer, G.~P. Salam, and G.~Soyez, {\it {The Lund Jet Plane}},
  \href{http://dx.doi.org/10.1007/JHEP12(2018)064}{JHEP {\bfseries 12} (2018)
  064}, \href{http://arxiv.org/abs/1807.04758}{{arXiv:1807.04758 [hep-ph]}}.

\bibitem{Carrazza:2019cnt}
S.~Carrazza and F.~A. Dreyer, {\it {Lund jet images from generative and
  cycle-consistent adversarial networks}},
  \href{http://dx.doi.org/10.1140/epjc/s10052-019-7501-1}{Eur. Phys. J.
  {\bfseries C79} (2019) 11, 979},
\href{http://arxiv.org/abs/1909.01359}{{arXiv:1909.01359 [hep-ph]}}.

\bibitem{Butter:2017cot}
A.~Butter, G.~Kasieczka, T.~Plehn, and M.~Russell, {\it {Deep-learned Top
  Tagging with a Lorentz Layer}},
  \href{http://dx.doi.org/10.21468/SciPostPhys.5.3.028}{SciPost Phys.
  {\bfseries 5} (2018) 3, 028},
  \href{http://arxiv.org/abs/1707.08966}{{arXiv:1707.08966 [hep-ph]}}.

\bibitem{Erdmann:2018shi}
M.~Erdmann, E.~Geiser, Y.~Rath, and M.~Rieger, {\it {Lorentz Boost Networks:
  Autonomous Physics-Inspired Feature Engineering}},
  \href{http://dx.doi.org/10.1088/1748-0221/14/06/P06006}{JINST {\bfseries 14}
  (2019) 06, P06006}, \href{http://arxiv.org/abs/1812.09722}{{arXiv:1812.09722
  [hep-ex]}}.

\bibitem{Bogatskiy:2020tje}
A.~Bogatskiy, B.~Anderson, J.~T. Offermann, M.~Roussi, D.~W. Miller, and
  R.~Kondor, {\it {Lorentz Group Equivariant Neural Network for Particle
  Physics}},  \href{http://arxiv.org/abs/2006.04780}{{arXiv:2006.04780
  [hep-ph]}}.

\bibitem{1811770}
X.~Ju and B.~Nachman, {\it {Supervised Jet Clustering with Graph Neural
  Networks for Lorentz Boosted Bosons}},
  \href{http://arxiv.org/abs/2008.06064}{{arXiv:2008.06064 [hep-ph]}}.

\bibitem{Shimmin:2021pkm}
C.~Shimmin, {\it {Particle Convolution for High Energy Physics}},
\newblock 7, 2021.
\newblock \href{http://arxiv.org/abs/2107.02908}{{arXiv:2107.02908 [hep-ph]}}.

\bibitem{Komiske:2017aww}
P.~T. Komiske, E.~M. Metodiev, and J.~Thaler, {\it {Energy flow polynomials: A
  complete linear basis for jet substructure}},
  \href{http://dx.doi.org/10.1007/JHEP04(2018)013}{JHEP {\bfseries 04} (2018)
  013}, \href{http://arxiv.org/abs/1712.07124}{{arXiv:1712.07124 [hep-ph]}}.

\bibitem{49372}
T.~Chen, S.~Kornblith, M.~Norouzi, and G.~E. Hinton, {\it A simple framework
  for contrastive learning of visual representations},
\newblock
\newblock \href{https://arxiv.org/abs/2002.05709}{2020}.

\bibitem{Mikuni:2020wpr}
V.~Mikuni and F.~Canelli, {\it {ABCNet: An attention-based method for particle
  tagging}},  \href{http://dx.doi.org/10.1140/epjp/s13360-020-00497-3}{Eur.
  Phys. J. Plus {\bfseries 135} (2020) 6, 463},
  \href{http://arxiv.org/abs/2001.05311}{{arXiv:2001.05311 [physics.data-an]}}.

\bibitem{Mikuni:2021pou}
V.~Mikuni and F.~Canelli, {\it {Point cloud transformers applied to collider
  physics}},  \href{http://dx.doi.org/10.1088/2632-2153/ac07f6}{Mach. Learn.
  Sci. Tech. {\bfseries 2} (2021) 3, 035027},
  \href{http://arxiv.org/abs/2102.05073}{{arXiv:2102.05073 [physics.data-an]}}.

\bibitem{Shmakov:2021qdz}
A.~Shmakov, M.~J. Fenton, T.-W. Ho, S.-C. Hsu, D.~Whiteson, and P.~Baldi, {\it
  {SPANet: Generalized Permutationless Set Assignment for Particle Physics
  using Symmetry Preserving Attention}},
  \href{http://arxiv.org/abs/2106.03898}{{arXiv:2106.03898 [hep-ex]}}.

\bibitem{Kasieczka:2019dbj}
A.~Butter {\em et al.}, {\it {The Machine Learning Landscape of Top Taggers}},
  \href{http://dx.doi.org/10.21468/SciPostPhys.7.1.014}{SciPost Phys.
  {\bfseries 7} (2019)  014},
  \href{http://arxiv.org/abs/1902.09914}{{arXiv:1902.09914 [hep-ph]}}.

\bibitem{benato2021shared}
L.~Benato, E.~Buhmann, M.~Erdmann, P.~Fackeldey, J.~Glombitza, N.~Hartmann,
  G.~Kasieczka, W.~Korcari, T.~Kuhr, J.~Steinheimer, H.~Stöcker, T.~Plehn, and
  K.~Zhou, {\it Shared data and algorithms for deep learning in fundamental
  physics},  \href{http://arxiv.org/abs/2107.00656}{{arXiv:2107.00656
  [cs.LG]}}.

\bibitem{Sjostrand:2014zea}
T.~Sj\"ostrand, S.~Ask, J.~R. Christiansen, R.~Corke, N.~Desai, P.~Ilten,
  S.~Mrenna, S.~Prestel, C.~O. Rasmussen, and P.~Z. Skands, {\it {An
  introduction to PYTHIA 8.2}},
  \href{http://dx.doi.org/10.1016/j.cpc.2015.01.024}{Comput. Phys. Commun.
  {\bfseries 191} (2015)  159},
  \href{http://arxiv.org/abs/1410.3012}{{arXiv:1410.3012 [hep-ph]}}.

\bibitem{deFavereau:2013fsa}
DELPHES 3, J.~de~Favereau, C.~Delaere, P.~Demin, A.~Giammanco,
  V.~Lema\^\i{}tre, A.~Mertens, and M.~Selvaggi, {\it {DELPHES 3, A modular
  framework for fast simulation of a generic collider experiment}},
  \href{http://dx.doi.org/10.1007/JHEP02(2014)057}{JHEP {\bfseries 02} (2014)
  057}, \href{http://arxiv.org/abs/1307.6346}{{arXiv:1307.6346 [hep-ex]}}.

\bibitem{Cacciari:2008gp}
M.~Cacciari, G.~P. Salam, and G.~Soyez, {\it {The anti-$k_t$ jet clustering
  algorithm}},  \href{http://dx.doi.org/10.1088/1126-6708/2008/04/063}{JHEP
  {\bfseries 04} (2008)  063},
  \href{http://arxiv.org/abs/0802.1189}{{arXiv:0802.1189 [hep-ph]}}.

\bibitem{Cacciari:2005hq}
M.~Cacciari and G.~P. Salam, {\it {Dispelling the $N^{3}$ myth for the $k_t$
  jet-finder}},  \href{http://dx.doi.org/10.1016/j.physletb.2006.08.037}{Phys.
  Lett. B {\bfseries 641} (2006)  57},
  \href{http://arxiv.org/abs/hep-ph/0512210}{{arXiv:hep-ph/0512210}}.

\bibitem{Cacciari:2011ma}
M.~Cacciari, G.~P. Salam, and G.~Soyez, {\it {FastJet User Manual}},
  \href{http://dx.doi.org/10.1140/epjc/s10052-012-1896-2}{Eur. Phys. J. C
  {\bfseries 72} (2012)  1896},
  \href{http://arxiv.org/abs/1111.6097}{{arXiv:1111.6097 [hep-ph]}}.

\bibitem{bahdanau2014neural}
D.~Bahdanau, K.~Cho, and Y.~Bengio, {\it Neural machine translation by jointly
  learning to align and translate},
  \href{http://arxiv.org/abs/1409.0473}{{arXiv:1409.0473 [cs.CL]}}.

\bibitem{luong2015effective}
M.-T. Luong, H.~Pham, and C.~D. Manning, {\it Effective approaches to
  attention-based neural machine translation},
  \href{http://arxiv.org/abs/1508.04025}{{arXiv:1508.04025 [cs.CL]}}.

\bibitem{vaswani2017attention}
A.~Vaswani, N.~Shazeer, N.~Parmar, J.~Uszkoreit, L.~Jones, A.~N. Gomez,
  L.~Kaiser, and I.~Polosukhin, {\it {Attention is all you need}},
  \href{http://arxiv.org/abs/1706.03762}{{arXiv:1706.03762 [cs.CL]}}.

\bibitem{ba2016layer}
J.~L. Ba, J.~R. Kiros, and G.~E. Hinton, {\it Layer normalization},
  \href{http://arxiv.org/abs/1607.06450}{{arXiv:1607.06450 [stat.ML]}}.

\bibitem{paszke2019pytorch}
A.~Paszke, S.~Gross, F.~Massa, A.~Lerer, J.~Bradbury, G.~Chanan, T.~Killeen,
  Z.~Lin, N.~Gimelshein, L.~Antiga, A.~Desmaison, A.~Köpf, E.~Yang, Z.~DeVito,
  M.~Raison, A.~Tejani, S.~Chilamkurthy, B.~Steiner, L.~Fang, J.~Bai, and
  S.~Chintala, {\it Pytorch: An imperative style, high-performance deep
  learning library},
\newblock \href{http://arxiv.org/abs/1912.01703}{2019}.

\bibitem{2020NumPy-Array}
C.~R. Harris, K.~J. Millman, S.~J. van~der Walt, R.~Gommers, P.~Virtanen,
  D.~Cournapeau, E.~Wieser, J.~Taylor, S.~Berg, N.~J. Smith, R.~Kern, M.~Picus,
  S.~Hoyer, M.~H. van Kerkwijk, M.~Brett, A.~Haldane, J.~Fernández~del Río,
  M.~Wiebe, P.~Peterson, P.~Gérard-Marchant, K.~Sheppard, T.~Reddy,
  W.~Weckesser, H.~Abbasi, C.~Gohlke, and T.~E. Oliphant, {\it Array
  programming with {NumPy}},
  \href{http://dx.doi.org/10.1038/s41586-020-2649-2}{Nature {\bfseries 585}
  (2020)  357–362}.

\bibitem{wang2021understanding}
F.~Wang and H.~Liu, {\it Understanding the behaviour of contrastive loss},
  \href{http://arxiv.org/abs/2012.09740}{{arXiv:2012.09740 [cs.LG]}}.

\end{thebibliography}\endgroup
\end{document}